\documentclass[12pt,amsonts,hyperref]{article}

\usepackage[left=1in,right=1in,top=1in,bottom=1in]{geometry}
\usepackage[utf8]{inputenc}
\usepackage{graphicx}
\usepackage{epsfig}
\usepackage{hyperref}
\usepackage{listings}
\usepackage{amsmath,amssymb}
\usepackage{caption}
\usepackage{subcaption}
\usepackage{color,soul}
\usepackage{authblk}
\usepackage{float}

\usepackage[square, numbers]{natbib}
\def\({\left(}
\def\){\right)}
\def\[{\left[}
\def\]{\right]}
\def\<{\langle}
\def\>{\rangle}
\def\tmp#1{}
\def\be{\begin{equation}}
\def\ee{\end{equation}}

\usepackage[table]{xcolor}
\usepackage{etoolbox}
\usepackage{pgf}
\definecolor{high}{HTML}{FF3333}
\definecolor{low}{HTML}{ECFF33}
\newcommand*{\opacity}{90}
\newcommand*{\minval}{1.0}
\newcommand*{\maxval}{5.7}

\newcommand*{\minvalmeanmed}{1.0}
\newcommand*{\maxvalmeanmed}{3.14}

\newcommand*{\minvaltaila}{0}
\newcommand*{\maxvaltaila}{49}

\newcommand*{\minvaltailb}{0}
\newcommand*{\maxvaltailb}{59.5}

\newcommand*{\minvaltailc}{0}
\newcommand*{\maxvaltailc}{80}

\newcommand*{\minvalmu}{0.1}
\newcommand*{\maxvalmu}{1.9}

\newcommand*{\minvalsigma}{0.6}
\newcommand*{\maxvalsigma}{1.3}

\newcommand*{\minvalemu}{0.0}
\newcommand*{\maxvalemu}{0.25}

\newcommand*{\minvalesigma}{0.0}
\newcommand*{\maxvalesigma}{0.15}

\newcommand*{\minvalcorr}{-1.0}
\newcommand*{\maxvalcorr}{+1.0}
\newcommand{\gradient}[1]{
    \ifdimcomp{#1pt}{>}{\maxval pt}{#1}{
        \ifdimcomp{#1pt}{<}{\minval pt}{#1}{
            \pgfmathparse{int(round(100*(#1/(\maxval-\minval))-(\minval*(100/(\maxval-\minval)))))}
            \xdef\tempa{\pgfmathresult}
            \cellcolor{high!\tempa!low!\opacity} #1
    }}
}

\newcommand{\gradienta}[1]{
    \ifdimcomp{#1pt}{>}{\maxvalmeanmed pt}{#1}{
        \ifdimcomp{#1pt}{<}{\minvalmeanmed pt}{#1}{
            \pgfmathparse{int(round(100*(#1/(\maxvalmeanmed-\minvalmeanmed))-(\minvalmeanmed*(100/(\maxvalmeanmed-\minvalmeanmed)))))}
            \xdef\tempa{\pgfmathresult}
            \cellcolor{high!\tempa!low!\opacity} #1
    }}
}

\newcommand{\taila}[1]{
    \ifdimcomp{#1pt}{>}{\maxvaltaila pt}{#1}{
        \ifdimcomp{#1pt}{<}{\minvaltaila pt}{#1}{
            \pgfmathparse{int(round(100*(#1/(\maxvaltaila-\minvaltaila)-(\minvaltaila*(100/(\maxvaltaila-\minvaltaila)))))}
            \xdef\tempa{\pgfmathresult}
            \cellcolor{high!\tempa!low!\opacity} #1
    }}
}

\newcommand{\tailb}[1]{
    \ifdimcomp{#1pt}{>}{\maxvaltailb pt}{#1}{
        \ifdimcomp{#1pt}{<}{\minvaltailb pt}{#1}{
            \pgfmathparse{int(round(100*(#1/(\maxvaltailb-\minvaltailb)-(\minvaltailb*(100/(\maxvaltailb-\minvaltailb)))))}
            \xdef\tempa{\pgfmathresult}
            \cellcolor{high!\tempa!low!\opacity} #1
    }}
}

\newcommand{\tailc}[1]{
    \ifdimcomp{#1pt}{>}{\maxvaltailc pt}{#1}{
        \ifdimcomp{#1pt}{<}{\minvaltailc pt}{#1}{
            \pgfmathparse{int(round(100*(#1/(\maxvaltailc-\minvaltailc)-(\minvaltailc*(100/(\maxvaltailc-\minvaltailc)))))}
            \xdef\tempa{\pgfmathresult}
            \cellcolor{high!\tempa!low!\opacity} #1
    }}
}

\newcommand{\colormu}[1]{
    \ifdimcomp{#1pt}{>}{\maxvalmu pt}{#1}{
        \ifdimcomp{#1pt}{<}{\minvalmu pt}{#1}{
            \pgfmathparse{int(100*(#1-\minvalmu)/(\maxvalmu-\minvalmu))}
            \xdef\tempa{\pgfmathresult}
            \cellcolor{high!\tempa!low!\opacity} #1
    }}
}

\newcommand{\colorsigma}[1]{
    \ifdimcomp{#1pt}{>}{\maxvalsigma pt}{#1}{
        \ifdimcomp{#1pt}{<}{\minvalsigma pt}{#1}{
            \pgfmathparse{int(100*(#1-\minvalsigma)/(\maxvalsigma-\minvalsigma))}
            \xdef\tempa{\pgfmathresult}
            \cellcolor{high!\tempa!low!\opacity} #1
    }}
}

\newcommand{\coloremu}[1]{
    \ifdimcomp{#1pt}{>}{\maxvalemu pt}{#1}{
        \ifdimcomp{#1pt}{<}{\minvalemu pt}{#1}{
            \pgfmathparse{int(100*(#1-\minvalemu)/(\maxvalemu-\minvalemu))}
            \xdef\tempa{\pgfmathresult}
            \cellcolor{high!\tempa!low!\opacity} #1
    }}
}

\newcommand{\coloresigma}[1]{
    \ifdimcomp{#1pt}{>}{\maxvalesigma pt}{#1}{
        \ifdimcomp{#1pt}{<}{\minvalesigma pt}{#1}{
            \pgfmathparse{int(100*(#1-\minvalesigma)/(\maxvalesigma-\minvalesigma))}
            \xdef\tempa{\pgfmathresult}
            \cellcolor{high!\tempa!low!\opacity} #1
    }}
}

\newcommand{\colorcorr}[1]{
    \ifdimcomp{#1pt}{>}{\maxvalcorr pt}{#1}{
        \ifdimcomp{#1pt}{<}{\minvalcorr pt}{#1}{
            \pgfmathparse{int(100*(#1-\minvalcorr)/(\maxvalcorr-\minvalcorr))}
            \xdef\tempa{\pgfmathresult}
            \cellcolor{high!\tempa!low!\opacity} #1
    }}
}
\begin{document}

\title{The impact of big winners on passive and active equity
investment strategies}

\author[]{Maxime Markov\thanks{Corresponding Author: \href{markov@theory.polytechnique.fr}{markov@theory.polytechnique.fr} } }
\author[]{Vladimir Markov}

\affil{}
\date{}

\date{}

\maketitle
\begin{abstract}

We investigate the impact of big winner stocks on the performance of active and passive investment strategies using a combination of numerical and analytical techniques. Our analysis is based on historical stock price data from 2006 to 2021 for a large variety of global indexes. We show that the log-normal distribution provides a reasonable fit for total returns for the majority of world stock indexes but highlight the limitations of this model. Using an analytical expression for a finite sum of log-normal random variables, we show that the typical return of a concentrated portfolio is less than that of an equally weighted index. This finding indicates that active managers face a significant risk of underperforming due to the potential for missing out on the substantial returns generated by big winner stocks. Our results suggest that passive investing strategies, that do not involve the selection of individual stocks, are likely to be more effective in achieving long-term financial goals.
\end{abstract}

\section{Introduction}

One of the most significant phenomena in the world of finance is the rise of passive investing. Active investing strategies give portfolio managers discretion to select individual securities, generally with the investment objective of outperforming a previously identified benchmark. In contrast, passive strategies use rule-based investing to track an index, typically by holding all its constituent assets or an automatically selected representative sample of those assets~\cite{Capocci:2001}. 

The share of passive investments is constantly increasing. For example, the rate of mutual and exchange-traded funds in the United States rose from 3\% in 1995 to 37\% in 2017~\cite{Anadu:2020}. Since then, the trend has only been upward, and passive investing is expected to overtake active investing by 2026~\cite{Seyffart:2021}. 
The reason for this shift is the persistent underperformance of active investing. Indeed, 99\% of actively managed US equity funds sold in Europe have failed to beat the S\&P 500 index over the period of 10 years from 2006 to 2016, while only two in every 100 global equity funds have outperformed the S\&P Global 1200. The situation is similar to active emerging market equity funds, 97\% of which underperform ~\cite{FT:2016,FT:2016b}.  
% original report  https://www.spglobal.com/spdji/en/documents/spiva/spiva-europe-year-end-2015.pdf
S\&P regularly publishes S\&P Indices Versus Active (SPIVA) research reports measuring the performance gap between actively managed and index funds~\cite{Spglobal:2021}.

Passive investing has several advantages over an active strategy.  First, passive investing products have lower fees relative to active mutual fund fees~\cite{Finrafee:2022}. The Morningstar investment research firm estimated that passive US fund investors saved \$38 billion  in fees in 2021 compared to what they would have paid to have their money in active funds. The higher fee effect is cumulative (this effect is also called "a tyranny of compounding costs") and represents a headwind for active investors. Second, active managers, like all humans, have cognitive and emotional biases. In particular, the disposition effect states that investors tend to sell winning investments and hold on to losing investments. Third, the impact of missing the market's best days can be huge. For example, missing the ten largest days in S\&P 500 leads to underperformance by 55\%, and seven of the best ten days occurred within two weeks of the ten worst days, which makes market timing challenging~\cite{JPM:2022}. Another important factor is the effect of a few big winner stocks that can grow by a factor of 10 or more over a long enough time-frame, which produces an outsized share of market returns. The last factor is the objective of this study.

In this paper, we explore the impact of big winners on investment performance from different perspectives for a wide variety of global indices. First, we examine the distribution of stock index returns using historical stock price data from 2006 to 2021 and quantify the difference between \textit{average} returns and \textit{typical} returns (approximated by a mode or median) for major stock indexes. We show that the log-normal distribution provides a reasonable fit for the total returns for most world stock indexes and highlight the limitations of this model. We use an analytical expression for the sum of $N$ log-normal random variables to quantify the ratio of the typical mean to the true mean as a function of the number of stocks in a portfolio. This shows how a typical small (concentrated) portfolio's performance differs from that of an index portfolio. 

Second, to better understand the mechanism of index returns, we fit a geometric Brownian motion (GBM) model to index constituents and extract index drift and volatility parameters. We observe a diverse range of relations between drift and volatility, which can be used to build a microscopic model of index returns, and quantify the effect of big winners. We also study a toy model with drift distributed according to a normal distribution and constant volatility. The ratios of mean to median and mean to mode are given by an analytical function of the parameters of the model.

\section{Empirical Data}

In this study, we use 16 years of yearly data from January 1, 2006, to December 31, 2021, with index constituents taken as of January 1, 2006. To avoid look-ahead bias, we take the index constituents as of the start date. We group indexes according to their geographical location into several groups, including the United States, Europe, Asia-Pacific (APAC), Japan, and BRIC (Brazil, India, and China) countries. We also study 10 sectors of S\&P 500 GICS Level 1 indexes to better understand the role of heterogeneity within the S\&P 500 index. The composition of the groups is given below:

\textbf{US indexes}: S\&P 500 (\textit{SPX}); NASDAQ Composite (\textit{CCMP}); Russell 3000 index, which is composed of 3,000 large US companies representing approximately 98\% of market capitalization of the investable US equity market (\textit{RAY}); Russell 2000 index, which consists of the smallest 2,000 companies in the Russell 3000 index representing approximately 8\% of the Russell 3000 index capitalization (\textit{RTY}); Russell 1000 index, which consists of the largest 1,000 companies in the Russell 3000 index (\textit{RIY}); Russell 1000 Value, which consists of Russel 1,000 companies with low price-to-book rations (\textit{RLV}); Russell 1000 Growth index with high price-to-book ratio (\textit{RLG}), and NASDAQ Biotechnology (\textit{NBI}).

\textbf{S\&P500 GICS Level 1 indexes}: Consumer Discretionary (\textit{S5COND}), Consumer Staples (\textit{S5CONS}), Energy (\textit{S5ENRS}), Financial (\textit{S5FINL}), Health Care (\textit{S5HLTH}), Information Technology (\textit{S5INFT}), Materials (\textit{S5MATR}), Communication Services (\textit{S5TELS}), Utilities (\textit{S5UTIL}), and Industrials (\textit{S5INDU}).

\textbf{European indexes} (including the UK): Deutsche Boerse German Stock Index (\textit{DAX}), French CAC 40 (\textit{CAC}), UK FTSE 100 (\textit{UKX}), Belgium BEL 20 (\textit{BEL20}), Spain IBEX 35 (\textit{IBEX}), Danish OMX Copenhagen 20 (\textit{KFX}), Swedish OMX Stockholm 30 index (\textit{OMX}), and Swiss Market Index (\textit{SMI}).

\textbf{APAC indexes}: Australia S\&P ASX 200 Index (\textit{AS51}).

\textbf{Japanese indexes}: Nikkei 225 (\textit{NKY}), Tokyo Price Index (\textit{TPX}).

\textbf{BRIC indexes}: Brazil Sao Paulo Stock Exchange Index (\textit{IBOV}), 
India NSE Nifty 40 Index (\textit{NIFTY}), MSCI India Index (\textit{MXIN}), Shanghai Stock Exchange Composite Index (\textit{SHCOMP}), and Shanghai Shenzhen CSI 300 Index (\textit{SHSZ300}).

In this study, we consider the performance of \textit{equally weighted} and \textit{fixed} portfolios or indexes only.  We neglect the effect of portfolio and index weights and rebalancing and concentrate on the impact that a few big-winner stocks have on a portfolio's long-term performance. In this framework,  investors allocate capital randomly and in equal units. It is a plausible model for uninformed investors. 

\section{Total Return Distribution}
\label{sec-return-distribution}

\begin{figure}[t]
  \includegraphics[width=0.5\linewidth]{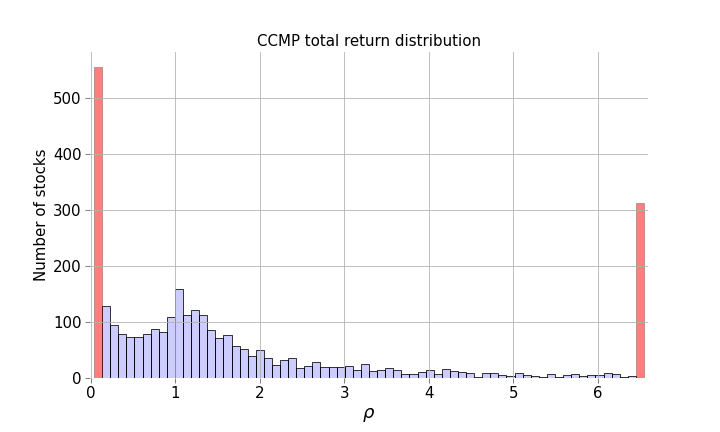}
  \includegraphics[width=0.5\linewidth]{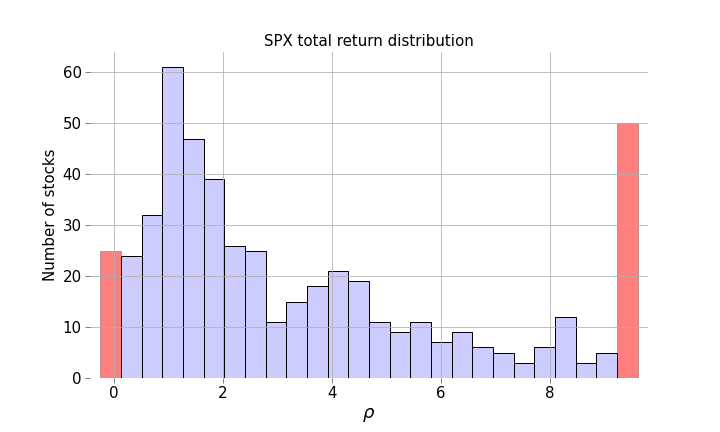}
  \caption{Total return histogram for the bimodal CCMP (left) and unimodal SPX (right) indexes. The histogram consists of the distribution body (blue bins), as well as the left and right tails grouped into single red bins. The left tail bin includes all stocks with $\ln(\rho) < -2$. The right tail bin takes in all stocks with the top 5\% returns.}
  \label{fig:histograms}
\end{figure}

We start by considering the distribution of the total return, defined as the ratio of the final price $X_T$  at time $t=T$ to the initial price $X_0$: $\rho = X_T/X_0$. To have positive support for $\rho$, we \textit{do not} subtract one in the definition of total return $\rho$. All prices are adjusted for dividends and splits. In Fig.~\ref{fig:histograms}, we show the total returns histogram for the CCMP (left panel) and SPX (right panel) indexes.  The histograms consist of the distribution body (blue bins), as well as the left and right cumulative bins highlighted in red. The left cumulative bin includes all beaten-down stocks satisfying condition $\ln(\rho)<-2$ (approximately 86\% loss). The right cumulative bin aggregates the best-performing stocks, with a total return of top 5\% in the index distribution. We use twice the number of bins determined by the Freedman-Diaconis~\cite{Freedman:1981} rule (see Table S1 and Figure S1 in Supplementary material~\footnote{\href{https://github.com/maxmarkov/stock-index/blob/master/suppmat/supplementary\%20passive\%20inversting.pdf}{Link to the supplementary material}}). 

Indexes can be divided into two groups: unimodal and bimodal. The first group includes indexes composed of stocks of well-established companies. The Belgium BEL20 and Swedish OMX indexes are typical examples from the first group. The left cumulative bin is small and fits well into the distribution body. In Section~\ref{section:macromodel}, we see that their distribution can be approximated by log-normal. Indexes belonging to the second group have excessive left cumulative bin, indicating a high number of depressed stocks that never recover. Examples are tech heavy NASDAQ (CCMP), biotech NBI indexes, and RAY, RTY, and AS51 indexes.

%The amplitude of the left cumulative bin is so %large that its removal is necessary to restore %log-normality.

In Table ~\ref{table1}, we present an analysis of the empirical distribution of the total index returns. First, we calculate the contribution to the (unweighted) index mean of the best 5, 10, and 25\% performing stocks, defined as follows:
\begin{equation}
\text{Top X \%} = 100 \% \cdot \left( \frac{E[\rho] - E[\rho|\rho<\rho_{top X \%}]}{E[\rho]} \right),
\end{equation}
where $E[\rho]=\frac{1}{N} \sum_{i=1}^N \rho_i$ and $E[\rho|\rho<\rho_{top X \%}]$ 
represents the mean total return for stocks that satisfy the condition $\rho<\rho_{top X \%}$. We highlight magnitude with colors ranging from yellow (small) to red (large). We find that the top 5\% of stocks in the CCMP, NBI, S5INFT, and AS51 indexes contribute over 40\% of the index's total return. Most European and Japanese indexes only have contributions of around 15\% and 20\%, respectively. Further, we compute the mean, median, and mode of the distribution, and the ratio of mean to median and mean to mode. The mode is calculated in a model-dependent way after fitting the empirical distribution with Gaussian kernel density. The mode estimation is not always stable for broad log-normal or close to exponential distributions. For unstable or bimodal distributions, we leave the mode column blank. The ratios show the difference between average returns and typical returns (approximated by a mode or median). Consistently, the highest ratios are found for indexes with the highest contribution from the top 5\% of stocks.

\begin{table}[!htbp]
\scalebox{0.95}{
{\setlength\tabcolsep{2.5pt}%
\begin{tabular}{|lr|rrr|rrl|cc|}
\multicolumn{10}{l}{16 years (between 2006-01-01 and 2021-12-31)} \\
 \hline
 index & $N$ & top 5\% & top 10\% & top 25\% &   mean &  median &  mode &  $\frac{mean}{median}$ &  $\frac{mean}{mode}$ \\
 \multicolumn{10}{l}{\textbf{US}} \\
\hline 
 SPX    &            498 &  \taila{25.9} & \tailb{35.0} & \tailc{54.1} &   4.26 &    2.37 &  1.37 & \gradienta{1.80} & \gradient{3.11} \\
 CCMP$^{*}$     &          3127 &  \taila{42.4} & \tailb{54.0} & \tailc{69.9} &   2.68 &    1.14 &  1.09$^{*}$ & \gradienta{2.36} & - \\
 RIY     &           974 &  \taila{25.0} & \tailb{35.2} & \tailc{55.3} &   3.97 &    2.06 &  1.33 & \gradienta{1.93} & \gradient{2.98} \\
 RTY$^{*}$      &          1981 &  \taila{36.4} & \tailb{48.4} & \tailc{65.9} &   3.07 &    1.35 &  1.17$^{*}$ & \gradienta{2.27} & - \\
 RAY$^{*}$     &           2974 &  \taila{31.9} & \tailb{43.6} & \tailc{62.9} &   3.36 &    1.55 &  1.21$^{*}$ & \gradienta{2.17} & - \\
 RLV     &           639 &  \taila{16.8} & \tailb{27.5} & \tailc{48.8} &   3.24 &    2.02 &  1.31 & \gradienta{1.61} & \gradient{2.47} \\
 RLG     &           636 &  \taila{26.6} & \tailb{36.9} & \tailc{57.0} &   4.54 &    2.27 &  1.34 & \gradienta{2.01} & \gradient{2.00} \\
 NBI$^{*}$      &           156 &  \taila{48.4} & \tailb{59.4} & \tailc{77.0} &   3.66 &    1.20 &  1.12$^{*}$ & \gradienta{3.04} & - \\
 \hline
 \multicolumn{10}{l}{\textbf{S\&P500}} \\
 \hline
 S5COND &             88  & \taila{32.5} & \tailb{42.6} & \tailc{63.2} &   4.94 &    2.01 &  1.38 & \gradienta{2.46} &       \gradient{3.58} \\
 S5CONS  &            38  & \taila{11.3}  & \tailb{18.6} & \tailc{33.9} &   4.37 &    4.18 &  4.29 & \gradienta{1.04} &       \gradient{1.01} \\
 S5ENRS  &            29  & \taila{11.4}  & \tailb{16.0} & \tailc{38.5} &   1.30 &    1.07 &  0.98 & \gradienta{1.21} &       \gradient{1.33} \\
 S5FINL$^{*}$  &            84  & \taila{18.6} & \tailb{28.5} & \tailc{49.2} &   2.34 &    1.54 &  1.32$^{*}$ & \gradienta{1.51} &      - \\
 S5HLTH  &            56  & \taila{13.7} & \tailb{21.9} & \tailc{38.3} &   4.25 &    3.13 &  1.48 & \gradienta{1.36} &       \gradient{2.87} \\
 S5INFT$^{*}$   &            78  & \taila{44.7} & \tailb{54.2} & \tailc{70.6} &   6.19 &    1.97 &  1.38 & \gradienta{3.14} &       - \\
 S5MATR  &            31  & \taila{11.0}  & \tailb{20.4} & \tailc{35.3} &   3.75 &    3.00 &  2.29 & \gradienta{1.25} &       \gradient{1.64} \\
 S5TELS  &             8  & \taila{22.4}  & \tailb{22.4}  & \tailc{34.8} &   1.60 &    1.47 &  1.50 & \gradienta{1.09} &       \gradient{1.07} \\
 S5UTIL  &            32  & \taila{14.6} & \tailb{22.1} & \tailc{33.7} &   3.40 &    2.89 &  1.81 & \gradienta{1.18} &       \gradient{1.88} \\
 S5INDU  &            53  & \taila{10.6}  & \tailb{19.2} & \tailc{34.7} &   6.16 &    4.77 &  3.52 & \gradienta{1.29} &       \gradient{1.75} \\
 \hline
 \multicolumn{10}{l}{\textbf{Europe}} \\
 \hline
 DAX     &            30  & \taila{11.8}  & \tailb{17.3} & \tailc{35.0} &   2.93 &    2.78 &  3.28 & \gradienta{1.05} & - \\
 CAC     &            40  & \taila{16.6} & \tailb{27.0} & \tailc{47.5} &   3.24 &    1.99  &  1.43 & \gradienta{1.62} & \gradient{2.27} \\
 UKX      &           102 & \taila{13.4} & \tailb{21.0} & \tailc{41.7} &   2.62 &    1.98  &  1.53 & \gradienta{1.32} & \gradient{1.71} \\
 BEL20   &            19  & \taila{18.5}  & \tailb{26.0} & \tailc{38.8} &   2.50 &    2.11 &  2.33 & \gradienta{1.18} & \gradient{1.07} \\
 IBEX   &             33  & \taila{16.8}  & \tailb{32.9} & \tailc{52.5} &   1.89 &    1.23 &  0.88 & \gradienta{1.53} & \gradient{2.15} \\
 KFX     &            20  & \taila{16.4} & \tailb{28.6} & \tailc{53.0} &   6.94 &    4.08  &  2.44 & \gradienta{1.70} & \gradient{2.84} \\
 OMX     &            30  & \taila{13.5}  & \tailb{18.6} & \tailc{39.6} &   6.00 &    4.91 &  3.52 & \gradienta{1.22} & \gradient{1.70} \\
 SMI     &            27  & \taila{22.6} & \tailb{28.1} & \tailc{45.8} &   2.87 &    1.88  &  0.99 & \gradienta{1.53} & \gradient{2.89} \\
 \hline
 \multicolumn{10}{l}{\textbf{APAC}} \\
 \hline
 AS51$^{*}$     &          200  & \taila{44.0} & \tailb{51.8} & \tailc{65.5} &   3.14 &    1.41 &  1.31$^{*}$ & \gradienta{2.23} & - \\
 \hline
  \multicolumn{10}{l}{\textbf{Japan}} \\
 \hline
 NKY     &           225 & \taila{17.8}  & \tailb{26.2} & \tailc{43.4} &   1.70 &    1.12 &  0.73 & \gradienta{1.51} & \gradient{2.33} \\
 TPX     &          1664 & \taila{19.8}  & \tailb{29.0} & \tailc{45.7} &   1.58 &    1.06 &  0.75 & \gradienta{1.48} & \gradient{2.11} \\
 \hline
  \multicolumn{10}{l}{\textbf{BRIC}} \\
 \hline
 IBOV     &             57 & \taila{17.7} & \tailb{26.9} & \tailc{48.6} &   4.01 &    2.51 &  1.58 & \gradienta{1.60} & \gradient{2.54} \\
 NIFTY    &             50 & \taila{12.8}  & \tailb{20.4} & \tailc{40.6} &   8.34 &    6.71 &  5.45 & \gradienta{1.24} & \gradient{1.53} \\
 MXIN    &              63 & \taila{20.0} & \tailb{28.2} & \tailc{44.4} &  10.12 &    7.31 &  5.72 & \gradienta{1.38} & \gradient{1.77} \\
 SHCOMP  &             873 & \taila{32.1} & \tailb{42.2} & \tailc{58.1} &   7.92 &    3.83 &  2.54 & \gradienta{2.06} & \gradient{3.11} \\
 SHSZ300$^{*}$ &             298 & \taila{34.1} & \tailb{44.9} & \tailc{59.6} &   6.87 &    3.26 &  2.43 & \gradienta{2.11} & - \\
\hline
\hline
\end{tabular}
}
}
\caption{Empirical analysis of the total return distribution. $N$ is the number of stocks; top X\% is the contribution to the index mean of the best X\% performing stocks; and the mean, median, and mode of index total return. Amplitudes are highlighted, with colors ranging from yellow to red. Bimodal indexes with a large cumulative left bin are marked with a star.} 
\label{table1}
\end{table}

\section{Mathematical Properties of Log-Normal Distribution}
\label{section-mathprop}

The potential returns of a stock depend on the time horizon.
While high-frequency returns (ranging from seconds to hours) exhibit a fat-tailed distribution with a power-law tail, longer-term returns (spanning quarters or years) tend to resemble a normal distribution more closely. The corresponding long-term total return $\rho$, which is  obtained by compounding and studied in this work can be modeled with the log-normal distribution. The log-normal distribution commonly appears as a basic model for multiplicative processes in biology, physics, or finance. 

A positive random variable $\rho$ is log-normally distributed if its natural logarithm is normally distributed:
\be
\ln ( \rho ) \sim Normal(\mu,\sigma^2),
\ee
where $\mu$ is the mean and $\sigma$ is the standard deviation of the corresponding normal distribution. 
Given these parameters, one can easily calculate the following distribution properties: mean = $e^{\mu+\sigma^2/2}$, median = $e^{\mu}$, and mode = $e^{\mu-\sigma^2}$.

The tail behavior is controlled by the shape parameter $\sigma$. 
As $\sigma$ increases, the log-normal distribution quickly becomes broad, with tail values much larger than the typical values from the distribution. In other words, a mode and a mean move away from each other, while a median stays in the same place. In the case of large $\sigma$, the distribution becomes concentrated around the mode, which is pushed to zero by the factor $e^{-\sigma^2}$. Therefore, the probability of drawing values much smaller than the median increases, and this results in many values much smaller than the median, all close to zero. Empirically, most indexes have $\sigma \approx1$. For $\sigma=1$, 69\% of the probability mass of log-normal distribution $LN(\mu,\sigma^2=1)$  lies below the average value. This corresponds to the odds of underperformance relative to an index for an uninformed stock picker.

Although many authors have focused on the difference between the median and the mean, we believe that the difference between the mode and the mean is more informative in the case of a wide-tailed log-normal distribution. The mode represents a typical return that a zero-intelligence investor can expect from the random stock selection process. The mean represents the index return of a passive investor.

\section{Macroscopic Model: Log-Normal Fit of Returns}
\label{section:macromodel}

To build a macroscopic model of index-normalized prices, we fit the logarithm of the total return distribution with normal, skew-normal, Laplace, and asymmetric Laplace distributions. The last two distributions allow us to study the effect of skewness and larger than normal tails. 

\begin{figure}[p]
  \includegraphics[width=0.5\linewidth]{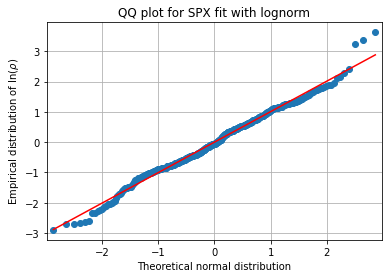}
  \includegraphics[width=0.5\linewidth]{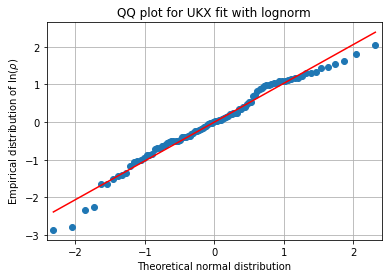}
  \includegraphics[width=0.5\linewidth]{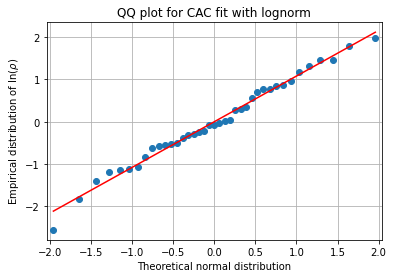}
  \includegraphics[width=0.5\linewidth]{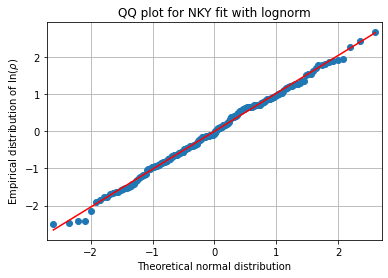}
  \includegraphics[width=0.5\linewidth]{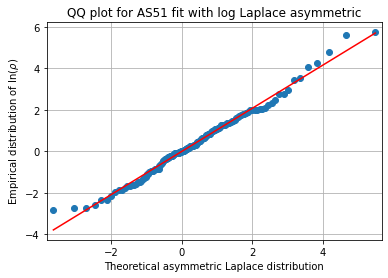}
  \includegraphics[width=0.5\linewidth]{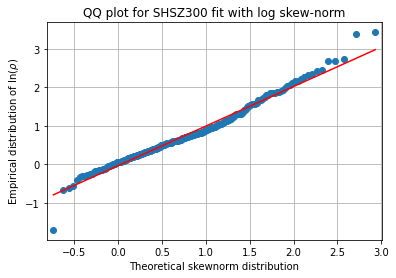}
  \caption{QQ plot of the total return distribution for several indexes. We fit log-normal (SPX, UKX, CAC, and NKY), logarithmic Laplace asymmetric (AS51), and logarithmic skew-normal (SHSZ300) distributions.}
  \label{fig:qqplot}
\end{figure}

As we saw in Section~\ref{sec-return-distribution}, some indexes have a bimodal distribution with a large left cumulative bin. This behavior is index idiosyncratic. 
To exclude this, we impose a total return threshold of $\ln(\rho)> -2$, which corresponds to the removal of stocks whose total return is $\rho< 0.14$ (about 86\% loss). After removing the left tail, the log-normal distribution fits most indexes well, except for a few cases when the right tail is better described by log-Laplace or logarithmic asymmetric Laplace distributions (such as AS51). The quality of fit can be assessed with a quantile-quantile (QQ) plot, which is shown for several indexes in Figure \ref{fig:qqplot}. If the fit and empirical distributions are linearly related, the points in the Q–Q plot will approximately lie on a line. If the points are above or below the line, the theoretical distribution is thinner or thicker than the empirical distribution. 
In Table~\ref{table2}, we show the parameters of the log-normal distribution ($\mu_{LN}$, $\sigma_{LN}$, and coefficient of variation $C=\sqrt{e^{\sigma_{LN}^2}-1}$) derived from the fits.

\begin{table}[p]
\scalebox{0.95}{
\begin{tabular}{|l|rr|rrr|rr|}
\multicolumn{8}{l}{16 years (between 2006-01-01 and 2021-12-31)} \\
\hline
 index &  $\mu_{LN}$ &   $\sigma_{LN}$ &  mean &  median &  mode &  $\sigma_{LN}^{2}$ &     C \\
\hline
\multicolumn{8}{l}{\textbf{US}} \\
\hline
SPX     & \colormu{0.95} & \colorsigma{1.02} &       4.37 &         2.60 &       0.92 &         1.04 &  1.35 \\
CCMP    & \colormu{0.41} & \colorsigma{1.10} &       2.78 &         1.51 &       0.45 &         1.22 &  1.54 \\
RIY     & \colormu{0.91} & \colorsigma{1.01} &       4.16 &         2.49 &       0.89 &         1.02 &  1.34 \\
RTY     & \colormu{0.59} & \colorsigma{1.07} &       3.19 &         1.80 &       0.58 &         1.14 &  1.46 \\
RAY     & \colormu{0.70} & \colorsigma{1.06} &       3.53 &         2.01 &       0.66 &         1.12 &  1.44 \\
RLV     & \colormu{0.84} & \colorsigma{0.93} &       3.59 &         2.32 &       0.97 &         0.87 &  1.18 \\
RLG     & \colormu{0.98} & \colorsigma{1.05} &       4.64 &         2.67 &       0.88 &         1.11 &  1.43 \\
NBI     & \colormu{0.60} & \colorsigma{1.27} &       4.05 &         1.81 &       0.36 &         1.60 &  1.99 \\
\hline
\multicolumn{8}{l}{\textbf{S\&P500}} \\
\hline
S5COND  & \colormu{0.98} & \colorsigma{1.15} &       5.18 &         2.67 &       0.71 &         1.33 &  1.66 \\
S5CONS  & \colormu{1.17} & \colorsigma{0.90} &       4.82 &         3.22 &       1.44 &         0.80 &  1.11 \\
S5ENRS  & \colormu{0.23} & \colorsigma{0.62} &       1.52 &         1.26 &       0.86 &         0.38 &  0.68 \\
S5FINL  & \colormu{0.60} & \colorsigma{0.99} &       2.95 &         1.81 &       0.69 &         0.97 &  1.28 \\
S5HLTH  & \colormu{1.10} & \colorsigma{0.84} &       4.29 &         3.01 &       1.48 &         0.71 &  1.02 \\
S5INFT  & \colormu{0.93} & \colorsigma{1.15} &       4.94 &         2.54 &       0.67 &         1.33 &  1.67 \\
S5MATR  & \colormu{1.07} & \colorsigma{0.74} &       3.85 &         2.92 &       1.68 &         0.55 &  0.86 \\
S5TELS  & \colormu{0.39} & \colorsigma{0.71} &       1.90 &         1.48 &       0.89 &         0.50 &  0.81 \\
S5UTIL  & \colormu{1.00} & \colorsigma{0.73} &       3.56 &         2.73 &       1.61 &         0.53 &  0.83 \\
S5INDU  & \colormu{1.46} & \colorsigma{0.99} &       6.97 &         4.29 &       1.63 &         0.97 &  1.28 \\
\hline
\multicolumn{8}{l}{\textbf{Europe}} \\
\hline
DAX     & \colormu{0.84} & \colorsigma{0.89} &       3.45 &         2.31 &       1.04 &         0.80 &  1.11 \\
CAC     & \colormu{0.76} & \colorsigma{0.97} &       3.43 &         2.15 &       0.84 &         0.94 &  1.25 \\
UKX     & \colormu{0.72} & \colorsigma{0.83} &       2.90 &         2.05 &       1.02 &         0.70 &  1.00 \\
BEL20   & \colormu{0.65} & \colorsigma{0.84} &       2.72 &         1.91 &       0.94 &         0.71 &  1.02 \\
IBEX    & \colormu{0.30} & \colorsigma{0.97} &       2.14 &         1.35 &       0.53 &         0.93 &  1.24 \\
KFX     & \colormu{1.61} & \colorsigma{0.94} &       7.77 &         4.98 &       2.05 &         0.89 &  1.20 \\
OMX     & \colormu{1.56} & \colorsigma{0.76} &       6.32 &         4.75 &       2.68 &         0.57 &  0.88 \\
SMI     & \colormu{0.59} & \colorsigma{1.10} &       3.29 &         1.80 &       0.54 &         1.20 &  1.53 \\
\hline
 \multicolumn{8}{l}{\textbf{APAC}} \\
\hline
AS51    & \colormu{0.57} & \colorsigma{1.00} &       2.90 &         1.77 &       0.66 &         0.99 &  1.30 \\
\hline
\multicolumn{8}{l}{\textbf{Japan}} \\
\hline
NKY     & \colormu{0.21} & \colorsigma{0.87} &       1.79 &         1.23 &       0.58 &         0.75 &  1.06 \\
TPX     & \colormu{0.11} & \colorsigma{0.86} &       1.63 &         1.12 &       0.53 &         0.75 &  1.05 \\
\hline
\multicolumn{8}{l}{\textbf{BRIC}} \\
\hline
IBOV    & \colormu{1.05} & \colorsigma{0.89} &       4.22 &         2.85 &       1.30 &         0.79 &  1.09 \\
NIFTY   & \colormu{1.65} & \colorsigma{1.23} &      11.12 &         5.22 &       1.15 &         1.51 &  1.88 \\
MXIN    & \colormu{1.88} & \colorsigma{1.12} &      12.21 &         6.54 &       1.88 &         1.25 &  1.58 \\
SHCOMP  & \colormu{1.47} & \colorsigma{1.00} &       7.17 &         4.33 &       1.59 &         1.01 &  1.32 \\
SHSZ300 & \colormu{1.34} & \colorsigma{0.93} &       5.92 &         3.83 &       1.60 &         0.87 &  1.18 \\
\hline
\hline
\end{tabular}
}
\caption{Parameters of log-normal distribution fit with condition $\ln \rho>-2$. $\mu_{LN}$ is location and $\sigma_{LN}$ is shape parameter; mean, median, and mode are calculated from known analytical expressions for log-normal distribution; and C is the coefficient of variation.}
\label{table2}
\end{table}

We also investigated indexes with constituents taken at the end of the period (2006-2021). 
The log-normal class described above fits them almost perfectly. This is consistent with the above statement, as the selected stocks always possess $\ln{\rho}>-2$ property.

\section{Typical Behavior of a Finite-Size Active Portfolio}
\label{section:finite-lognorm-sampling}

Thus far, we have investigated the distribution of returns, quantifying the difference between average and typical returns for major stock indexes. In our analysis, a typical return (mode or median) indirectly represents an active investment strategy. In turn, a direct way to mimic an active manager's portfolio selection is to randomly pick a finite number of stocks from the distribution. 
To simulate the process, we form a portfolio of $N$ stocks by drawing $N$ random numbers from the log-normal distribution and summing them to obtain the aggregate performance. The parameters of the log-normal distribution for each index have been derived in Section~\ref{section:macromodel} and summarized in Table~\ref{table2}.

Let us define the mean $R_N$ of $N$ total  returns $\rho_i=X^i_T/X_0^i$ at time $T$ as follows:
\be
R_N=\frac{1}{N}\sum_{i=1}^N \rho_i
\ee
There are two limiting scenarios for the behavior of $R_N$. First, it is equal to the true average value $\langle \rho \rangle$ if the distribution is narrow or if $N$ is asymptotically large:
\be
R_N=\langle \rho \rangle
\label{eq_average}
\ee
On the other hand, $R_N$ can deviate strongly from Eq.~\ref{eq_average} for broad distributions. In this case, its behavior is defined by a few $k$ largest terms:
\be
R_N \simeq \frac{1}{N} \sum_k max_k(\rho_1,...,\rho_n)
\ee

According to Romeo \textit{et al.}~\cite{Romeo:2003}, the sum of log-normal variables has three regimes: narrow (I), moderately broad (II), and very broad (III). Depending on the variance $\sigma_{LN}^2$ of the log-normal distribution, the typical sample mean $R_N$ (\textit{i.e.}, the mode of the mean distribution) is given by:
\begin{equation}
\begin{aligned}
  \text{Regime I: } \quad  & \sigma_{LN}^2 \ll 1 \quad && R_{N,I} = e^{\mu_{LN}}  \\
  \text{Regime II: } \quad & \sigma_{LN}^2 \lesssim 1 \quad && R_{N,II} = \langle \rho \rangle  \left( 1 + \frac{C^2}{N} \right)^{-3/2},\ C=\sqrt{e^{\sigma_{LN}^2}-1} \\
  \text{Regime III: } \quad & \sigma_{LN}^2 \gg 1 \quad && R_{N,III} = \langle \rho \rangle  \exp \left[ -\frac{3}{2} \, 
\frac{\sigma_{LN}^2}{N^{\ln(3/2)/\ln2}} \right],
\end{aligned}
\label{eq_sm}
\end{equation}
where $N$ is the number of random variables $\rho$ drawn from log-normal distribution, $\mu_{LN}$ and $\sigma_{LN}^2$ are the distribution parameters, and $\langle \rho \rangle$ is the true mean.

\begin{figure}[t]
  \centering
  \includegraphics[width=0.8\linewidth]{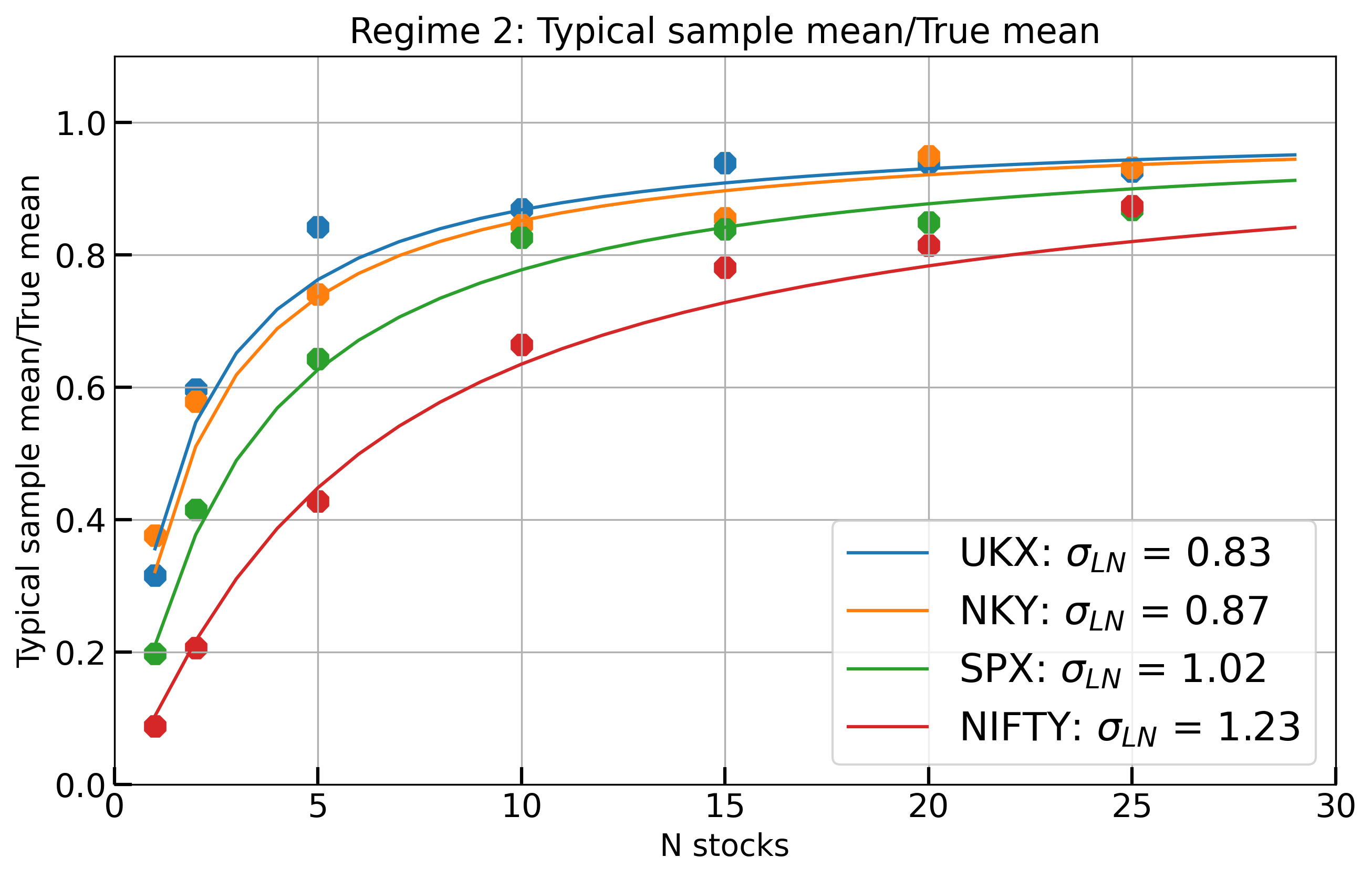}
  \caption{The ratio of the typical sample mean (active portfolio) to the true mean (passive portfolio) as a function of the number of stocks $N$ for several selected indexes in Regime II. The results of the analytical approach and Monte Carlo simulation are shown by solid curves and dots, respectively. The indexes are arranged according to their log-normal distribution variances $\sigma_{LN}$.}
  \label{fig:regime2}
\end{figure}

As can be seen from the variance values $\sigma_{LN}^{2}$ in Table~\ref{table2}, we always have a moderately broad log-normal distribution corresponding to Regime II. Figure \ref{fig:regime2} shows the ratio of the typical sample mean (active portfolio) to the true mean (passive portfolio) as a function of the number of stocks $N$ for several selected indexes. The results obtained using the analytical approach of Eq.~\ref{eq_sm} correspond well to the results of the Monte-Carlo simulations (dots). Indexes with smaller variance (UKX and NKY) require fewer stocks in a portfolio than indexes with greater variance (NIFTY and NBI) to achieve the performance of the passive strategy. Nevertheless, even smaller variance indexes require quite many of stocks ($N > 20$) for this.

\section{Microscopic Models of Indexes: Geometric Brownian Motion Fit of Index Constituents}

In this section, we study the distribution of drift and volatility of constituents for each index and build a microscopic model of index returns.

\subsection{Geometric Brownian motion model of a single stock}

It is often assumed that the price of a stock is described by a stochastic process $X_t$, which follows a GBM and satisfies the following stochastic differential equation:
\be
\frac{dX_{t+1}}{X_t} = \mu \, dt + \sigma\,dW_t,
\ee
where $W_t$ is a Wiener process with percentage drift $\mu$ and percentage volatility $\sigma$. The solution of this equation is well known and is given by:
\be
X_t=X_0 e^{\mu-\frac{1}{2} \sigma^2 T} e^{\sigma W_t},
\ee
where $X_0$ is the stock starting price. Thus, a stock has the expected price at time $T$:
\be
E[X_T]=X_0 e^{\mu T}, \,\,\, Var[X_T]=X_0^2 e^{2 \mu T} (e^{\sigma^2 T}-1)
\ee

We use maximum likelihood estimation of the GBM percentage drift $\hat{\mu}$ and percentage volatility $\hat{\sigma}$ parameters:
\be
\hat \mu=\frac{1}{T}\ln\left(\frac{X_T}{X_0}\right)+\frac{1}{2} \hat \sigma^2
\label{eq:likelihood-drift}
\ee
\be
\hat \sigma^2=\frac{1}{T}\left(\sum_{t=1}^{T} \ln^2\left(\frac{X_{t}}{X_{t-1}}\right)-\frac{\ln^2\left(\frac{X_T}{X_0}\right)}{(T-1)}\right),
\label{eq:likelihood-volatility}
\ee
where we sum over 15 yearly periods from 2006 to 2021.

\begin{figure}
  \includegraphics[width=0.5\linewidth]{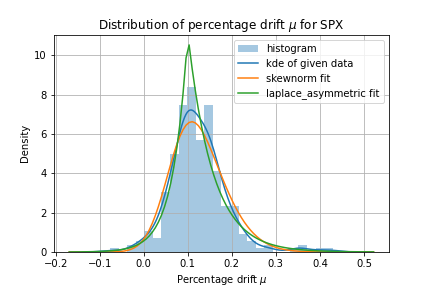}
  \includegraphics[width=0.5\linewidth]{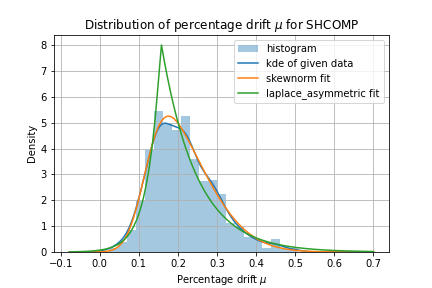}
  \includegraphics[width=0.5\linewidth]{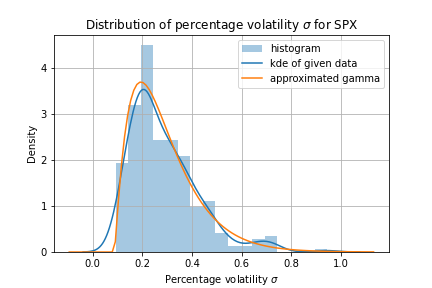}
  \includegraphics[width=0.5\linewidth]{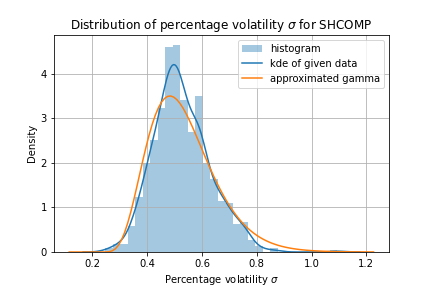}
  \includegraphics[width=0.5\linewidth]{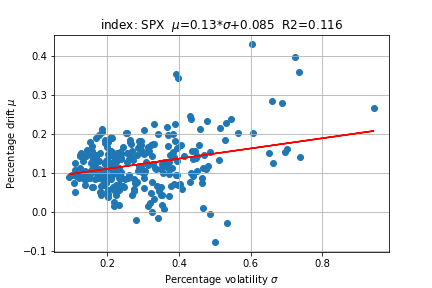}
  \includegraphics[width=0.5\linewidth]{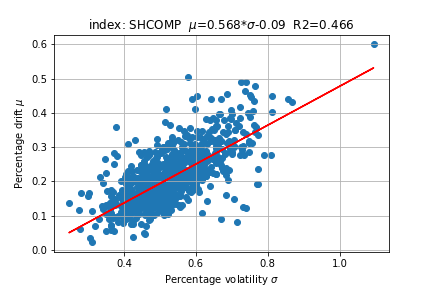}
\caption{Distribution of drift  $\hat\mu$  and volatility $\hat\sigma$  parameters for SPX and SHCOMP indexes}
  \label{fig:gbm_dist}
\end{figure}

Figure~\ref{fig:gbm_dist} shows the distribution of estimated drift $\hat{\mu}$ and volatility $\hat{\sigma}$ parameters for SPX and SHCOMP indexes. A reasonable fit of the drift distribution is given by the skewed-normal distribution (top panels), while volatility is well approximated (for practical purposes, middle panels) by a gamma distribution.  The correlations between $\hat{\mu}$ and  $\hat{\sigma}$ are mostly positive but market-dependent. The relationship quantifies the efficient market hypothesis assumption that higher risks correlate with higher potential returns. The fitted parameters and other statistics are summarized in Table~\ref{table3}. 

The delisted stocks are not traded during the whole period, making parameter estimations unreliable. Thus, we exclude them from the analysis. %The delisted stocks were excluded from this analysis. 

\begin{table*}[p]%\centering
\scalebox{0.90}{
\begin{tabular}{|l|rrrrrrrrrrrc|}
\hline
index  &  E[$\hat{\mu}$] &  std($\hat{\mu}$) &  $\mu_{\zeta}$ &  $\mu_{\omega}$ & $\mu_{\alpha}$  &  E[$\hat{\sigma}$] &  $\sigma_\alpha$ &  $\sigma_\beta$ &     $a$ &     $b$ &    $R^2$ &  corr[$\hat{\mu},\hat{\sigma}$] \\
\hline
\hline
\multicolumn{13}{l}{\textbf{US}} \\
\hline
SPX    & \coloremu{0.12} & \coloresigma{0.06} & 0.06 & 0.09 & 1.88 & 0.29 & 2.15 & 10.70 & 0.13 & 0.09 & 0.12 & \colorcorr{0.35} \\
CCMP   & \coloremu{0.15} & \coloresigma{0.14} & 0.02 & 0.20 & 1.91 & 0.42 & 2.12 &  6.62 & 0.28 & 0.03 & 0.29 & \colorcorr{0.55} \\
RIY    & \coloremu{0.13} & \coloresigma{0.07} & 0.05 & 0.11 & 3.60 & 0.31 & 2.35 & 10.12 & 0.17 & 0.08 & 0.25 & \colorcorr{0.52} \\
RTY    & \coloremu{0.14} & \coloresigma{0.10} & 0.04 & 0.14 & 1.65 & 0.38 & 2.27 &  8.11 & 0.26 & 0.04 & 0.26 & \colorcorr{0.51} \\
RAY    & \coloremu{0.14} & \coloresigma{0.09} & 0.05 & 0.12 & 1.76 & 0.35 & 2.50 &  9.16 & 0.21 & 0.06 & 0.24 & \colorcorr{0.50} \\
RLV    & \coloremu{0.12} & \coloresigma{0.06} & 0.06 & 0.09 & 2.95 & 0.30 & 1.64 &  8.11 & 0.16 & 0.07 & 0.29 & \colorcorr{0.55} \\
RLG    & \coloremu{0.14} & \coloresigma{0.06} & 0.07 & 0.09 & 2.59 & 0.30 & 3.03 & 12.81 & 0.14 & 0.09 & 0.13 & \colorcorr{0.37} \\
NBI    & \coloremu{0.15} & \coloresigma{0.15} & 0.32 & 0.23 &-2.98 & 0.51 & 1.99 &  5.31 & 0.21 & 0.08 & 0.06 & \colorcorr{0.30} \\
\hline
\multicolumn{13}{l}{\textbf{S\&P500}} \\
\hline
S5COND & \coloremu{0.14} & \coloresigma{0.07} & 0.07 & 0.10 & 1.98 & 0.32 & 2.71 & 11.79 &  0.14 &  0.10 &  0.13 & \colorcorr{0.38} \\
S5CONS & \coloremu{0.11} & \coloresigma{0.04} & 0.07 & 0.05 & 2.27 & 0.18 & 0.34 &  5.13 &  0.39 &  0.04 &  0.27 & \colorcorr{0.54} \\
S5ENRS & \coloremu{0.06} & \coloresigma{0.05} & 0.06 & 0.05 & 0.00 & 0.37 & 5.26 & 20.00 & -0.13 &  0.11 &  0.11 & \colorcorr{-0.39} \\
S5FINL & \coloremu{0.10} & \coloresigma{0.04} & 0.06 & 0.06 & 1.68 & 0.32 & 2.10 &  9.23 &  0.10 &  0.07 &  0.16 & \colorcorr{0.41} \\
S5HLTH & \coloremu{0.13} & \coloresigma{0.05} & 0.07 & 0.08 & 2.84 & 0.23 & 3.08 & 19.17 &  0.35 &  0.05 &  0.45 & \colorcorr{0.68} \\
S5INFT & \coloremu{0.17} & \coloresigma{0.07} & 0.09 & 0.11 & 3.31 & 0.33 & 1.76 &  9.67 &  0.14 &  0.11 &  0.14 & \colorcorr{0.45} \\
S5MATR & \coloremu{0.15} & \coloresigma{0.07} & 0.07 & 0.11 & 2.76 & 0.37 & 1.52 &  6.13 &  0.24 &  0.06 &  0.37 & \colorcorr{0.61} \\
S5TELS & \coloremu{0.05} & \coloresigma{0.02} & 0.04 & 0.02 & 0.50 & 0.17 & 0.07 &  1.13 & -0.45 &  0.12 &  1.00 & \colorcorr{-1.00} \\
S5UTIL & \coloremu{0.09} & \coloresigma{0.03} & 0.12 & 0.05 &-1.59 & 0.18 & 1.48 & 20.00 & -0.23 &  0.12 &  0.20 & \colorcorr{ -0.47} \\
S5INDU & \coloremu{0.14} & \coloresigma{0.05} & 0.19 & 0.07 &-1.92 & 0.27 & 3.21 & 20.00 &  0.12 &  0.11 &  0.01 & \colorcorr{0.16} \\
\hline
\multicolumn{13}{l}{\textbf{Europe}} \\
\hline
DAX    & \coloremu{0.09} & \coloresigma{0.08} & 0.02 & 0.10 & 1.54 & 0.32 & 4.51 & 17.34 &  0.06 &  0.07 &  0.05 & \colorcorr{0.29} \\
CAC    & \coloremu{0.09} & \coloresigma{0.05} & 0.03 & 0.07 & 2.13 & 0.27 & 1.32 &  9.51 &  0.06 &  0.07 &  0.02 & \colorcorr{0.15} \\
UKX    & \coloremu{0.09} & \coloresigma{0.06} & 0.01 & 0.10 & 2.90 & 0.28 & 1.37 &  8.34 &  0.12 &  0.05 &  0.11 & \colorcorr{0.34} \\
BEL20  & \coloremu{0.12} & \coloresigma{0.08} & 0.02 & 0.13 & 5.63 & 0.36 & 0.27 &  1.06 &  0.33 & -0.00 &  0.69 & \colorcorr{0.83} \\
IBEX   & \coloremu{0.04} & \coloresigma{0.08} & 0.11 & 0.11 &-1.53 & 0.31 & 1.48 &  8.33 & -0.20 &  0.09 &  0.04 & \colorcorr{-0.22} \\
KFX    & \coloremu{0.15} & \coloresigma{0.07} & 0.06 & 0.11 & 4.08 & 0.34 & 0.29 &  1.59 &  0.18 &  0.09 &  0.13 & \colorcorr{0.37} \\
OMX    & \coloremu{0.11} & \coloresigma{0.06} & 0.18 & 0.09 &-5.89 & 0.27 & 0.83 &  7.64 & -0.15 &  0.16 &  0.30 & \colorcorr{-0.60} \\
SMI    & \coloremu{0.07} & \coloresigma{0.06} & 0.13 & 0.09 &-1.77 & 0.27 & 0.31 &  2.09 & -0.15 &  0.11 &  0.06 & \colorcorr{-0.26} \\
\hline
\multicolumn{12}{l}{\textbf{AS51}} \\
\hline
AS51   & \coloremu{0.11} & \coloresigma{0.08} & 0.02 & 0.12 & 2.11 & 0.34 & 1.51 & 6.94 &  0.18 &  0.04 &  0.14 & \colorcorr{0.37} \\
\hline
\multicolumn{13}{l}{\textbf{Japan}} \\
\hline
NKY    & \coloremu{0.07} & \coloresigma{0.06} & 0.02 & 0.07 & 1.11 & 0.31 & 4.46 & 20.00 & 0.20 &  0.01 &  0.13 & \colorcorr{0.37} \\
TPX    & \coloremu{0.07} & \coloresigma{0.06} & 0.01 & 0.08 & 1.53 & 0.29 & 4.94 & 20.00 & 0.26 & -0.01 &  0.24 & \colorcorr{0.49} \\
\hline
\multicolumn{13}{l}{\textbf{BRIC}} \\
\hline
IBOV   & \coloremu{0.15} & \coloresigma{0.06} & 0.21 & 0.09 &-1.83 & 0.40 & 0.41 & 2.24 &  0.09 &  0.12 & -0.14 & \colorcorr{-0.16} \\
NIFTY  & \coloremu{0.19} & \coloresigma{0.07} & 0.13 & 0.09 & 1.28 & 0.40 & 3.88 &12.03 &  0.29 &  0.08 &  0.42 & \colorcorr{0.65} \\
MXIN   & \coloremu{0.20} & \coloresigma{0.07} & 0.13 & 0.10 & 1.67 & 0.41 & 3.66 &11.24 &  0.30 &  0.08 &  0.43 & \colorcorr{0.66} \\
SHCOMP & \coloremu{0.21} & \coloresigma{0.08} & 0.12 & 0.13 & 3.23 & 0.53 & 6.05 & 20.00 &  0.57 & -0.09 &  0.47 & \colorcorr{0.68} \\
SHSZ300 & \coloremu{0.20} & \coloresigma{0.07} & 0.11 & 0.11 & 3.08 & 0.52 & 6.81 &20.00 &  0.54 & -0.09 &  0.52 & \colorcorr{0.72} \\
\hline
\end{tabular}
}
\caption{Analysis of fitted drift and volatility distributions. E[$\hat \mu$] is the mean of MLE of the drift $\hat \mu$, std($\hat \mu$) is the standard deviation of $\hat \mu$, parameters of skew-normal fit of $\hat \mu$ distribution are  location $\mu_{\zeta}$ , scale $\mu_{\omega}$ and shape $\mu_{\alpha}$; E[$\hat \sigma$] is the mean of estimated volatility $\hat \sigma$, parameters of gamma distribution fit of $\hat \sigma$ distribution are shape $\sigma_\alpha$ and inverse scale(rate) $\sigma_\beta$; parameters of robust (Huber) linear regression $\hat{\mu}=a\hat{\sigma}+b$ and corresponding $R^2$; correlation coefficient between drift and volatility corr[$\hat{\mu},\hat{\sigma}$].}
\label{table3}
\end{table*}

\subsection{Analytically solvable model of a stock index: Effect of distributed drift}

Let us introduce an analytically solvable toy model for index returns. Assume that each stock price $X^i$ in an index follows a geometric Brownian motion: 
\be
\frac{dX^i_{t+1}}{X^i_t} = \mu_i\, dt + \sigma\,dW_t,
\ee
with the percentage drift distributed according to the normal distribution $\mu_i\sim N(\mu_d,\sigma_d)$. Equivalently, the drift can be written as follows:
\be
\mu_i=\mu_d+\sigma_d Z,
\ee
where $Z \sim N(0,1)$. The GBM solution over the time period $t\in [0,T]$ is given by the standard integration and the Ito formula:
\be
X_T=X_0 e^{\mu_d-\frac{1}{2} \sigma^2 T} e^{\sigma W_t +\sigma_d T Z}
\ee
where $W_t=N(0,T \sigma^2)$.
The last term can be simplified as follows:
\be
\sigma W_t +\sigma_d T Z=N(0,T \sigma^2)+N(0,T^2 \sigma_d^2)=N(0,\sigma^2 T+\sigma_d^2 T^2)
\label{sumwz}
\ee
Thus, the final stock price $X^i_T$ of $i$ stock in an index is as follows:
\begin{equation}
X^i_T=X_0 e^{\mu_d T-\frac{1}{2}\sigma^2 T + \sqrt{\sigma^2 T+\sigma_d^2 T^2}Z}, 
\label{indexlogn}
\end{equation}
where we neglected the covariance between stocks.

From Eq.~\ref{indexlogn}, one can see that the total return of a randomly chosen stock at time $t=0$ follows a log-normal distribution with $\mu_m=\mu_d T -\frac{1}{2}\sigma^2 T$ and  $\sigma_m^2=\sigma^2 T+\sigma_d^2 T^2$. %\hl{In the model with normally distributed drift, both stock and index total returns follow log-normal distribution.} 
The mean, median, and mode of the distribution are given by:
\be
\ln{(mean)}=\mu_m+\frac{\sigma_m^2}{2} = \mu_d T+\frac{1}{2} \sigma_d^2T^2
\ee
\be
\ln{(median)}=\mu_m=\mu_d T-\frac{1}{2}\sigma^2T
\ee
\be
\ln{(mode)}=\mu_m-\sigma_m^2=\mu_d T-\frac{3}{2} \sigma^2 T - \sigma_d^2T^2
\ee
Over time $T$, more than half of all stocks in the index underperform the index return by a factor of: 
\be
\frac{mean}{median} = \exp\left[\frac{1}{2}\sigma^2 T+\frac{1}{2}\sigma_d^2T^2\right]
\ee
and a typical stock underperforming the index by factor 
\be
\frac{mean}{mode} = \exp\left[\frac{3}{2}\sigma^2T+\frac{3}{2}\sigma_d^2T^2\right]
\ee

The variance component $\sigma^2_d T^2$ of the drift plays an important role. It mathematically represents the effect of the continuous compounding of winners pushing the average return up, while a large body of distribution is concentrated around the mode and goes to zero. Thus, the log-normal model  generates a small number of extreme winners and a large number of stocks with drifts centered around mode $\exp[\mu_{m}-\sigma_{m}^{2}]$.

The model can be matched against empirical data. For the SPX index, we have $\mu_d=0.12$, $\sigma_d=0.03$, and mode of volatility $\sigma=0.1$. The model result for mean/median is 1.26 vs. the empirical value of 1.42. %For consistency, we exclude delisted stocks here. 
We note that non-Gaussianity (skewness or fatter tails) of the drift distribution and correlation between the drift $\mu$ and volatility $\sigma$ should be considered to improve the microscopical model of index returns. 

\section{Conclusions}

In this work, we study the impact of big winner stocks on the behavior of equally weighted indexes and concentrated portfolios. We observe that long-term total returns obtained by compounding of simple returns exhibit skewness for all major stock indexes. The degree of skewness is index specific. We find that the top 5\% of stocks in the CCMP, NBI, S5INFT, and AS51 indexes contribute over 40\% of the index’s total return. Most European and Japanese indexes only have contributions of around 15\% and 20\%, respectively. The highest mean-to-mode ratios are found for indexes with the highest contribution from the top 5\% of stocks. Some indexes have a bimodal distribution with a large left cumulative bin, which is index idiosyncratic. After removing the left bin, the log-normal distribution fits most indexes well. The key observation made through our analysis, utilizing an analytical expression for a finite sum of log-normal random variables, is that the typical return of a concentrated portfolio is typically smaller than that of an equally weighted index. We also fit the historical stock returns of the indexes with the GBM. A reasonable fit of the drift distribution is given by the skewed-normal distribution, while volatility is well approximated by a gamma distribution. In a toy model of index returns with normally distributed drift, we show how the parameters of the distribution control the mean-to-mode ratio of index returns.

\section*{Acknowledgements}
The author gratefully thanks Olga Markova and Vasilisa Markov for their helpful discussions and suggestions.

%%% All references mybiblio.bib file %%%
%\bibliography{mybiblio}  

\end{document}